# The electronic and transport properties of a molecular junction studied by an integrated piecewise thermal equilibrium approach

M. -H. Tsai[*], T. –H. Lu and Y. –H. Tang[**]

*Department of Physics, National Sun Yat-Sen University, Kaohsiung 80424 Taiwan*

An integrated piecewise thermal equilibrium approach based on the first-principles calculation method has been developed to calculate bias dependent electronic structures and current- and differential conductance-voltage characteristics of the gold-benzene-1,4-dithiol-gold molecular junction. The calculated currents and differential conductance have the same order of magnitude as experimental ones. An electron transfer was found between the two electrodes when a bias is applied, which renders the two electrodes to have different local electronic structures. It was also found that when Au 5d electrons were treated as core electrons the calculated currents were overestimated, which can be understood as an underestimate of the Au-S covalent bonding and consequently the contact potential barrier and the replacement of delocalized Au 5d carriers by more itinerant delocalized Au 6sp carriers in the electrodes.

## I. INTRODUCTION

Studies of electron transport through organic molecules as the possible components in molecular electronics are both of fundamental and technological importance. Fabrication and measurements of current-voltage (I-V) and/or conductance-voltage (C-V) characteristics of metal-molecule-metal systems using various organic molecules have been achieved by Reed et al. [1], Chen et al. [2], Tian et al. [3], Xiao, Xu, and Tao [4], Dadosh et al. [5], and Kiguchi et al. [6] Among the various organic molecules considered the benzene-1,4-dithiol (BDT) molecule is the simplest. The experimentally observed transport properties of these metal-molecule-metal systems have attracted intensive theoretical investigations [7-39]. Nitzan and Ratner have reviewed the research of the electron transport in molecular wire junctions [40].

Theoretical calculations of the electric current, $I$, have been commonly based on the standard expression

$$I = \frac{2e}{\hbar}\int_{-\infty}^{\infty}T(E)[f(E-\mu_1)-f(E-\mu_2)]dE \quad (1),$$

where $f$ is the Fermi-Dirac distribution function, $\mu_1$ and $\mu_2$ are the bias dependent chemical potentials at the two metal electrodes and $T(E)$ is the transmission coefficient obtained by, for examples, the scattering theory of transport [41] and non-equilibrium Green function (NEGF) approach [42-46]. The NEGF formalism of Keldysh, Kadanorf and Baym has usually been employed in conjunction with the density-functional-theory (DFT) based first-principles calculation methods to calculate transport properties of molecular junctions. For examples, Xue and Ratner [28] carried out a microscopic study of end group effect on electrical transport through individual molecules, Kondo et al. [29] studied the contact-structure dependence of transport properties of a single BDT molecule between Au(111) electrodes, Chen et al. [30] studied the control of substituent ligand over current through molecular devices, and Waldron et al. [31] studied nonlinear spin current and magnetoresistance of a Ni-BDT-Ni molecular magnetic tunnel junction (MTJ). On the other hand, Geng et al. [32] used the Lippmann-Schwinger scattering method in conjunction with DFT studied the impact of metal electrode and molecular orientation on the conductance of a single BDT molecule.

In the scattering theory and NEGF approach, the molecular junction is usually divided into three parts, namely two semi-infinite electrodes and an "extended molecule", which includes BDT and a part of the electrode in contact or bonded with the molecule. The two semi-infinite electrodes were regarded as reservoirs of itinerant carriers. In the NEGF/DFT approach, the effect of bias is manifested in the external potential in the Kohn-Sham equation. The molecule-electrode interaction is manifested as the self-energy. Brandbyge et al. [44] found a correspondence between the transfer matrices derived from the scattering theory and NEGF. Faleev et al. [33] found that the use of NEGF reduced the calculated current by about one half from that obtained by a zero-bias transmission coefficient at a bias of 4 Volts for the gold-BDT-gold molecular junction.

In previous theoretical studies, the molecule is usually regarded as chemisorbed at the hollow site on the flat surfaces of the Au electrodes. In this chemisorption picture, the S atom is bonded with multiple Au atoms. Despite intensive efforts of previous theoretical investigations, the calculated currents for the gold-BDT-gold junction were persistently one to three orders of magnitude larger than the experimental data of Reed et al., except the ones by Wang and Lou [15] and Wang, Fu and Lou [16], who used the hybrid density function theory of quantum chemistry. When Di Ventra et al. [24] considered that the S end of BDT is attached to the flat jellium surface, the calculated currents were about 500 times larger than experimental data of Reed et al. [1] However, when they considered the case that each S end is bonded with a single Au atom and the single Au atom is then attached to the flat jellium surface, the calculated currents were greatly improved and reduced down to only about one order of magnitude larger than experimental data. A similar contact geometry, in which each S end of BDT is bonded with a single Au atom, was also considered by Cuevas et al. [35]

Guided by the finding of Di Ventra et al. [24], the geometry of the molecule-electrode contact in this study is chosen such that each S end of BDT is bonded with a single Au atom and this single Au atom is then chemisorbed at the three-fold hollow site of the Au(111) surface. To avoid the complexity of dividing the molecular junction into three parts, namely two electrodes and an extended molecule, and corresponding division of interactions, potentials and Hamiltonian matrix, an integrated piecewise thermal equilibrium approach has been developed and tested on the gold-BDT-gold molecular junction. The calculated currents and conductance have the same order of magnitudes as those obtained by Reed et al. For example, the current per molecule obtained in this study is 0.050μA at a bias of 1.5Volts vs. ~0.03μA obtained by Reed et al. at the same bias. And zero-bias conductance obtained in this study is 0.0062μS, while the experimental value of Reed et al. was ~0.002μS [1].

**II. The structural model and the first-principles methods used**

There are two kinds of structural models used for studying surfaces. One is the semi-infinite model, which contains only one surface. Another is the slab model, which includes the single slab model with (in principle) semi-infinite vacuum regions on both sides and infinitely repeated-slab model or supercell model usually used in pseudopotential calculations. The slab model has been widely employed in first-principles studies of surface science. In this study the single slab model is chosen for the gold electrode. Due to the screening effect of conduction electrons, a three-atomic-layer slab already has a total density of states, D(E), similar to that of the bulk crystal of metals. Thus, a three-layer Au(111) slab model is chosen. An extension to a thicker-slab model can be considered later. A schematic drawing of the optimized Au-atom-BDT-Au-atom molecule and the gold slabs is shown in Fig. 1. The Au atom bonded with the S atom is located at the three-fold hollow site above the Au(111) film surface. The plane of the benzene ring is parallel with the [1,0] basis vector of the Au(111) film. A (3x2) Au(111) periodicity is chosen for the self-assembled monolayer (SAM) of BDT molecules, so that the unit cell has totally 38 Au, 2 S, 6 C and 4 H atoms. This unit cell is chosen to avoid direct overlapping among BDT molecules. The periodicity of the Au(111) film is 2.885 Å, so the two-dimensional unit cell is 8.655 Å x 4.997 Å. The atomic positions of the BDT molecule have been optimized by the first-principles molecular dynamics (MD) calculation method [47,48]. The MD method is based on the real-space norm-conserving pseudopotential method [49,50] implemented with the Ceperley-Alder exchange-correlation potentials.[51] The basis set is composed of Bloch sums of $s$, $p_x$, $p_y$ and $p_z$ pseudo-atom orbitals. This method efficiently calculates the forces on atoms using the modified Hellmann-Feynman theorem [52], which are then used to predict the equilibrium atomic positions iteratively until the magnitudes of the forces are less than 0.1 eV/Å. The modified pseudofunction (PSF) method [53,54] implemented with the linear theory of Andersen [55] is used to calculate the electronic structure and the total energy. The basis set used in the PSF method contains two sets of Bloch sums of muffin-tin orbitals with exponentially decaying spherical Hankel functions, which describe localized lower-energy states, and oscillating Neumann tailing functions, which describe higher-energy conduction-bandstates. Inside the muffin-tin sphere, the muffin-tin orbital is a linear combination of the solution of the atomic Kohn-Sham equation and its energy derivative within the linear theory of Andersen. The muffin-tin orbital and its radial derivative are continuous at the muffin-tin radius. The pseudofunctions are constructed by replacing muffin-tin-sphere parts of the muffin-tin orbitals by smooth mathematical functions, so that their Bloch sums are expandable by a small number of plane waves; they are simply devised to calculate the interstitial and non-spherical parts of the Hamiltonian matrix elements efficiently using the fast-Fourier-transform (FFT) technique. The PSF method uses the Hedin-Lundqvist [56] form of local density approximation for the exchange-correlation potential. The four special $\vec{k}$ point scheme of Cunningham [57] is used to obtain self-consistent charge densities and potentials.

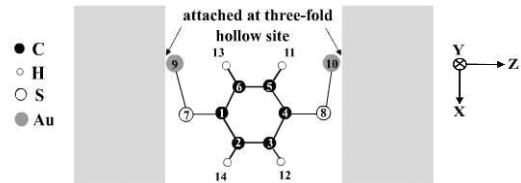

Fig. 1 A schematic drawing of the gold-SAM-gold tri-layer system

The average C-C, C-H and C-S bond lengths obtained in this study are 1.428Å, 1.159Å and 1.929Å, respectively, which differ from the experimental C-C and C-H bond lengths in a benzene molecule of 1.399Å and 1.101Å [58] by 2.07% and 5.26% , respectively, and the C-S single-bond length in $(CH_2)_2S(CH_2)_2$ of 1.839Å [58] by 4.89%. The C-S bond lengths obtained by Faleev et al. [33] and Stokbro et al. [19] were 1.74Å and 1.75Å, respectively. The calculation of the total energies with respect to the adjustment of the Au-S distance and the Au-S-C angle yields a Au-S bond length of 2.20Å, which is about 7% smaller than 2.42Å [12], 2.39Å[19], and 2.37Å [35] obtained previously for the S atom coordinated with multiple Au atoms, and an Au-S-C bond angle of 98.9°, which is within the range of known bond angles of S, for example : $\angle F-S-F = 98.2°$ in $SF_2$, $\angle Cl-S-Cl = 103°$ in $SCl_2$ and $\angle C-S-H = 103°$ in $CH_3CH_5-SH$ [58]. Chen et al. [30] obtained an Au-S bond length of 2.37Å and an Au-S-C bond angle of 102.8° for the Au-atom-BDT-Au-atom molecule. In comparison, Yourdshahyan, Zhang and Rappe obtained a much larger Au-S-C bond angle between 132° and 138° for the long-chain alkane thiols $[CH_3(CH_2)_{n-1}SH]$ chemisorbed on the Au(111) surface [59]. The calculated partial densities of states (PDOS) of S and C atoms and the Au atoms that are bonded with the S atoms are shown in Fig. 2, in which the Fermi level, $E_F$, is chosen as the zero energy. Fig. 2 shows that there is a band gap of 0.77eV and that the highest occupied molecular orbital (HOMO) energy levels are composed of S 3p, C 2p and Au 5d hybridized states, while the lowest unoccupied molecular orbital (LUMO) energy levels are composed of S 3p, C 2p, Au 6sp and 5d hybridized states, . The wide spreading of the Au 5d states and the overlapping of PDOS's of Au 5d and S 3p states indicate a significant part of covalent bonding between Au and S via Au 5d orbitals.

**III. The integrated piecewise thermal equilibrium approach**

In this approach, the non-equilibrium distribution function of electrons in the metal-SAM-metal tri-layer system with a bias applied to the two metal electrodes is approximated by a piecewise thermal equilibrium distribution function. This is an adequate approximation because electrons can response fast enough to the local potential similar to the success of the local density approximation (LDA) for the exchange-correlation energy, which approximates the non-uniform electron density as piecewise uniformly distributed. With this approximation the non-equilibrium distribution function of electrons is given as

$$f(E,T,\mu(z)) = [1 + \exp(\{E - \mu(z)\}/\{k_B T\})]^{-1} ,$$

where $E$, $T$, $\mu(z)$ and $k_B$ are the eigenenergy, absolute temperature, local chemical potential and Boltzmann constant, respectively. The z coordinate is perpendicular to the tri-layer system. Since the resistivity of the molecule is much higher than that of the electrodes, the potential drop due to the bias is expected to be entirely in the molecule as already shown by Kondo et al. [29] and Faleev et al. [33].Thus, the z-dependent local chemical potential is approximated as

$$\mu(z) = \begin{cases} \frac{\mu_1+\mu_2}{2} + \frac{(\mu_2-\mu_1)}{2}\cos\left[\frac{\pi(z-b+2d)}{2(d-b)}\right] & -d < z < -b \\ \mu_1 & -b \leq z \leq -t \\ \mu_1 + \frac{(\mu_2-\mu_1)}{2t}(z+t) & -t < z < t \\ \mu_2 & t \leq z \leq b \\ \frac{\mu_1+\mu_2}{2} + \frac{(\mu_2-\mu_1)}{2}\cos\left[\frac{\pi(z-b)}{2(d-b)}\right] & b < z \leq d \end{cases} \quad (2),$$

where $\mu_1$ and $\mu_2$ are the constant chemical potentials in the two metal slabs within (-b,-t) and (t,b), respectively, and $\mu(z)$ drops linearly from $\mu_1$ to $\mu_2$ in the Au-atom-BDT-Au-atom molecule between $-t$ and $t$. The bias voltage, $V_{bias}$, is related to $\mu_1$ and $\mu_2$ by $(-e)V_{bias} = \mu_2 - \mu_1$. The parts of $\mu(z)$ outside the tri-layer system between (-d, -b) and (b,d) are smooth mathematical functions extended beyond z=-b and z=b with $\mu(-d)=\mu(d)$, which are chosen such that $\mu(z)$ is continuous and differentiable at z=±b, so that $\mu(z)$ can be accurately expanded by a minimal number of plane waves with wave numbers of $2\pi m/d$, where $m$ is an integer, in order to be easily implemented in the PSF method. It has been tested that the minimal plane wave expansion of $\mu(z)$ is very accurate if $d$ is sufficiently larger than $b$. In this study, $\mu(z)$ is fixed, which can be determined self-consistently later. The external potential, $v_{ext}(z)$, added to the effective potential in the Kohn-Sham equation is given by $v_{ext}(z) = \mu(z) - \Delta v_{coul}$. The Coulomb potential difference across the tri-layer system, $\Delta v_{coul}$, is calculated as $\Delta v_{coul} = <v_{coul}(x,y,b)> - <v_{coul}(x,y,-b)>$, where $< >$ stands for the average over the (x,y) plane and $v_{coul}(x,y,z)$ is the Coulomb potential calculated from the total electron charge density and ion core charges. Both $v_{ext}(z)$ and $v_{coul}(x,y,z)$ are determined self-consistently. The subtraction of $\Delta v_{coul}$ ensures that the difference of the total Coulomb potential of the system, $v_{ext}(z)+v_{coul}(x,y,z)$, between the outside surfaces of the two slabs is equal to $(-e)V_{bias}$.

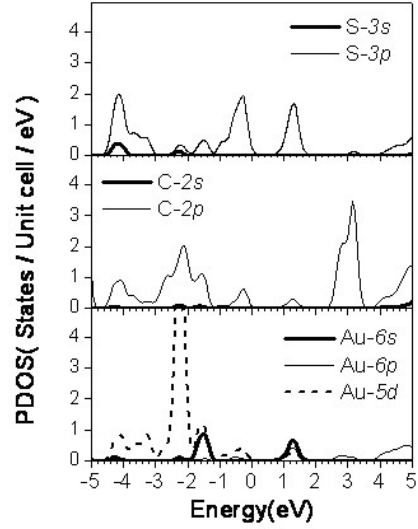

Fig. 2 PDOS's of the SAM of Au-BDT-Au molecules (without Au electrodes). The peak value of the Au 5d feature located at -2.3eV is 8.0 states/unit cell/eV.

If $E_{n,\bar{k}}$ and $\Psi_{n,\bar{k}}(x,y,z)$ are the *n*th eigenenergy and eigenfunction of the Kohn-Sham equation of the whole tri-layer system under the external potential, $v_{ext}(z)$, the occupation number of this state, $f_{n,\bar{k}}$, is calculated as the expectation value of the Fermi-Dirac distribution operator, i.e.

$$f_{n,\bar{k}} = \langle \Psi_{n,\bar{k}} | f(E_{n,\bar{k}}, T, \mu(z)) | \Psi_{n,\bar{k}} \rangle \quad (3)$$

The electron density, $n(x,y,z)$, is then calculated as

$$n(x,y,z) = 2\sum_{\bar{k}} \omega_{\bar{k}} \sum_n f_{n,\bar{k}} \Psi^*_{n,\bar{k}}(x,y,z) \Psi_{n,\bar{k}}(x,y,z) \quad (4)$$

The factor 2 is due to spin degeneracy and $\omega_{\bar{k}}$ is the weighting factor of the special $\bar{k}$ point. As usual, $E_{n,\bar{k}}$ and $\Psi_{n,\bar{k}}(x,y,z)$ as well as $n(x,y,z)$ are determined self-consistently. Within the PSF method, the Bloch sums of pseudofunctions, $P_l^{\bar{k}}(x,y,z)$, are identical to the basis wavefunctions, i.e. Bloch sums of muffin-tin orbitals, in the interstitial region and are expanded as

$$P_l^{\bar{k}}(x,y,z) = \sum_{\bar{G}_{//}} \sum_{G_z} P_{\bar{G}_{//},G_z,l}^{\bar{k}} \exp\{-i(\bar{G}_{//}-\bar{k})\cdot(x\hat{i}+y\hat{j})\}\exp\{-iG_z z\} \quad (5),$$

where $\bar{G}_{//}$ is the two-dimensional reciprocal vectors and $G_z = 2\pi m/d$ with $m=0,\pm1,\pm2,\pm3,......$ and *l* is the index of the basis wavefunction in the basis set. The Bloch sums of pseudofunctions can be readily decomposed into +z- and –z-direction traveling waves. The two +z and –z traveling waves, $P_{l,+}^{\bar{k}}(x,y,z)$ and $P_{l,-}^{\bar{k}}(x,y,z)$, are constructed as

$$P_{l,+}^{\bar{k}}(x,y,z) = \tfrac{1}{\sqrt{2}} \sum_{\bar{G}_{//}} \sum_{G_z<0} P_{\bar{G}_{//},G_z,l}^{\bar{k}} \exp\{-i(\bar{G}_{//}-\bar{k})\cdot(x\hat{i}+y\hat{j})\}\exp\{-iG_z z\}$$

$$P_{l,-}^{\bar{k}}(x,y,z) = \tfrac{1}{\sqrt{2}} \sum_{\bar{G}_{//}} \sum_{G_z>0} P_{\bar{G}_{//},G_z,l}^{\bar{k}} \exp\{-i(\bar{G}_{//}-\bar{k})\cdot(x\hat{i}+y\hat{j})\}\exp\{-iG_z z\} \quad (6),$$

respectively. The factor $1/\sqrt{2}$ is to compensate the spit of one state, which is occupied with two electrons, into two states occupied with four electrons. Note that each Bloch sum is associated with a crystal momentum, $\hbar\bar{k}$, where $\bar{k}$ is a wavevector in the two-dimensional Brillouin zone. Thus, basis wavefunctions are traveling waves in the (x,y) plane. However, they are not traveling waves in the z-direction. The decomposition of basis wavefunctions or Bloch sums of pseudofunctions in the z direction is similar to the decomposition of standing waves confined in a square box, say between z= – b and z=b, into +z- and –z-direction traveling waves. For plane-wave based first-principles methods such as the pseudopotential and linear augmented plane wave methods, the basis wavefunctions are already traveling waves, so that decomposition is not needed.

If the eigenvector corresponding to the eigenvalue $E_{n,\bar{k}}$ is $\{a_{n,l}^{\bar{k}}\}$, the +z and –z traveling waves in the interstitial region are

$$\Psi_{n+}^{\bar{k}}(x,y,z) = \tfrac{1}{\sqrt{2}} \sum_l a_{n,l}^{\bar{k}} \sum_{\bar{G}_{//}} \sum_{G_z<0} P_{\bar{G}_{//},G_z,l}^{\bar{k}} \exp\{-i(\bar{G}_{//}-\bar{k})\cdot(x\hat{i}+y\hat{j})\}\exp\{-iG_z z\}$$

$$\Psi_{n-}^{\bar{k}}(x,y,z) = \tfrac{1}{\sqrt{2}} \sum_l a_{n,l}^{\bar{k}} \sum_{\bar{G}_{//}} \sum_{G_z>0} P_{\bar{G}_{//},G_z,l}^{\bar{k}} \exp\{-i(\bar{G}_{//}-\bar{k})\cdot(x\hat{i}+y\hat{j})\}\exp\{-iG_z z\} \quad (7)$$

The total current densities in the interstitial region moving in +z and –z directions, $j_\pm$, can be expressed as

$$j_\pm(x,y,z) = \frac{ie\hbar}{m_e} \sum_{\bar{k}} \omega_{\bar{k}} \sum_n f_{n,\bar{k}} \left[ \Psi_{n\pm}^{\bar{k}*} \left(\frac{\partial}{\partial z}\Psi_{n\pm}^{\bar{k}}\right) - \left(\frac{\partial}{\partial z}\Psi_{n\pm}^{\bar{k}*}\right) \Psi_{n\pm}^{\bar{k}} \right] \quad (8)$$

by applying the standard quantum mechanical current density equation, $\bar{j} = \frac{ie\hbar}{2m_e}[\Psi^*(\nabla\Psi) - (\nabla\Psi^*)\Psi]$, for the electron with a charge $(-e)$, where $m_e$ is the electron mass. These two current densities can be efficiently calculated by FFT and expressed as

$$j_\pm(x,y,z) = \sum_{\bar{G}_{//}} \sum_{G_z} J_{\bar{G}_{//},G_z,\pm} \exp\{-i\bar{G}_{//}\cdot(x\hat{i}+y\hat{j})\}\exp\{-iG_z z\} \quad (9)$$

The (x,y) plane averaged current densities are given by

$$j_\pm(z) = \sum_{G_z} J_{0,G_z,\pm} \exp\{-iG_z z\} \quad (10)$$

The current through the molecule is calculated as $I = A[j_+(z_m) + j_-(-z_m)]$, where *A* is the area of the unit cell, which contains one molecule, and $z_m$ is the end position of the molecule, at which pseudofunctions are identical to real basis wave functions in $z=\pm z_m$ (x,y) planes. Since the tri-layer structural model is not a closed circuit, the current $I' = A[j_-(z_m) + j_+(-z_m)]$ compensate *I*, so that there is no net current flow. However, in the experimental current measurements, a closed circuit has to be connected. For example, the two gold electrodes are connected to a battery or a power supply. The current *I'* will be drained into the battery or power supply. Then, there will be a net current, *I*, flows through the molecule. The present approach mimics the drift current, which is the difference between two opposite-direction currents due to traveling waves moving in opposite directions, in a finite-sized section of metal.

The traveling wave formalism is conceptually the same as the calculation of the transmission probability of a particle passing through a one-dimensional (1-D) potential barrier/well given in most quantum mechanics textbooks. The $Aj_+(z_m)$ and $Aj_-(-z_m)$ currents are analogous to $\frac{2e}{\hbar}\int_{-\infty}^{\infty} T(E)f(E-\mu_1)dE$ and $-\frac{2e}{\hbar}\int_{-\infty}^{\infty} T(E)f(E-\mu_2)dE$ currents given in equation (1). The former corresponds to the current due to traveling waves which impinge from the left (–z) side and pass through the potential barrier/well or the molecule to reach the right (+z) side. The latter corresponds to the current in opposite direction and the net current is the difference between them. The major difference between present formalism and those based on the scattering theory and NEGF approach is that in the present case the +z-direction (-z-direction) traveling waves at the +z (-z) end of the molecule are solved directly from the Hamiltonian of the whole system just like the transmitted traveling waves solved in the 1-D potential barrier/well problem given in standard quantum mechanics textbooks. In contrast, the whole electrode-molecule-electrode system is divided into two electrodes and one extended-molecule parts and $T(E)$ is calculated using the Lippmann-Schwinger equation or the correlation (lesser Green) function in previous calculations.

The difference with the 1-D potential barrier/well case is that in the present case the slabs are terminated by vacuum. The effect of the slab-vacuum interface can be reduced by considering a thicker-slab model. Recently, Zhang et al. [60] embedded an absorbing layer in the slab of an infinitely repeated slab-molecule-slab-vacuum supercell model and derived a three-dimensional generalized Bloch theorem to calculate quantum transport of a molecular junction. The absorbing layer was devised to simulate a reservoir, which drains the current. However, the current in a closed electric circuit does not flow in this way. For any section in a closed circuit, if the current drains from one end of this section, in the steady state there must be a current source which supplies the same current to the other end of this section. Thus the absorption-layer model is not considered in the present study.

## III. Electronic structures of the gold-SAM-gold tri-layer system

Figure 3 shows the calculated PDOS's of (a) Au atoms and (b) S, C, and H atoms of the gold-SAM-gold tri-layer system at $V_{bias}=0$. The calculated PDOS's of the bulk gold crystal is also shown in the top panel of Fig. 3(a) to elucidate the contact effect and to show the adequacy of the three-layer slab model for the electrode. In these figures, the constant chemical potential or Fermi level, $E_F$, is chosen as zero energy. The top panel shows that the leading edge of the major Au 5d features is located at ~1.0 eV below $E_F$. In the vicinity of $E_F$, the states are mainly Au 6sp states. The Au 6s and 6p bands spread over a very large energy range, which indicates that delocalized Au 6s and 6p electrons are highly mobile. The Au 5d band spreads with a band width of about 6.5 eV. For the $V_{bias}=0$ gold-SAM-gold tri-layer system, the overall width of the major 5d features of the Au atoms in the two slabs is about 6.5 eV, which is the same as the 5d band width of the bulk gold crystal. However, the PDOS's of these Au 5d bands don't drop to zero at –1.0eV, but tail off gradually up to about 8eV above $E_F$, which indicates that a small percentage of Au 5d states are highly delocalized. The overall width of the major 5d features of the single Au atom bonded with the S atom of BDT is about 6.0eV, which is narrower than those of Au atoms in the slabs. The major 5d features of this single Au atom shifts to higher energies with the leading edge extending above $E_F$ due to bonding with the S atom. Similar to the bulk gold crystal, the Au 6s and 6p bands spread over a very large energy region. In the vicinity of $E_F$, PDOS's of Au 6s, 6p and 5d states are approximately equal. The result that the overall width of the major Au 5d features in the central layer (denoted by Au2l in Fig. 3(a)) of the slab is the same as that of the bulk gold crystal and the shapes of 5d, 6s and 6p bands roughly resemble those of the bulk gold crystal indicates that the three-layer slab model considered in this study for the gold electrode is adequate. The PDOS's of S and C atoms in BDT shown in Fig. 3(b) differ greatly from those of the Au-atom-BDT-Au-atom molecule shown in Fig. 2, which has an energy gap of 0.77eV with HOMO and LUMO bands dominated by S 3p states similar to previous calculations for the BDT molecule [29]. However, S 3p band is greatly broadened when BDT is attached to the Au slabs and the energy gap disappears. The major features in the PDOS's of S, C and H states at $V_{bias}=1.5$ Volts are similar to those shown in Fig. 3(b) except some fine details.

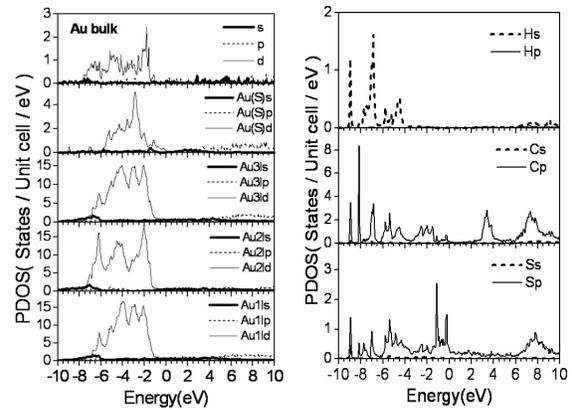

Fig. 3 (a) PDOS's of Au atoms in $1^{st}$-, $2^{nd}$-, $3^{rd}$-layers and the one bonded with the S atom denoted as Au(S) and (b) of S, C and H in the zero-bias gold-SAM-gold tri-layer system.

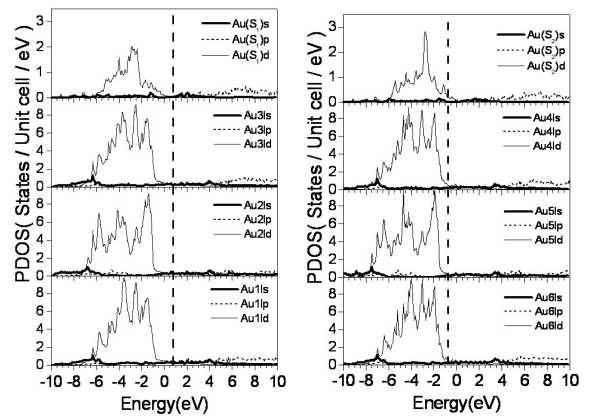

Fig. 4 PDOS's of Au atoms in (a) $1^{st}$-$3^{rd}$ layers and the one bonded with the S atom in slab 1, Au($S_1$), and (b) $4^{th}$-$6^{th}$ layers and the one bonded with the S atom in slab 2, Au($S_2$). The vertical dash lines mark local chemical potentials.

The PDOS's of Au 5d bands for $V_{bias}=1.5$ Volts are shown in Figs 4(a) and (b), in which the zero energy is chosen as the local chemical potential at the center of the tri-layer system. The leading edges of the major 5d features of the Au atoms in slabs 1 and 2 with $\mu_1$=+0.75 eV and $\mu_2$=−0.75 eV shift downward from $\mu_1$ and upward toward $\mu_2$, respectively, which indicates an electron transfer from slab 1 to slab 2 that lowers and raises local electrostatic potential in slab 1 and 2, respectively, when a bias is applied. This effect causes the density of states at $\mu_1$, $D(\mu_1)$, in slab 1 to be different from the corresponding $D(\mu_2)$ in slab 2 in contrast to previous calculations, which usually assumed $D(\mu_1)=D(\mu_2)$.

**V. The transport properties of the gold-SAM-gold tri-layer system**

Figures 5(a) and (b) shows the current-voltage (I-V) and differential conductance-voltage (C-V) characteristics of the gold-SAM-gold tri-layer system up to $V_{bias}$=1.5Volts. The calculated currents and differential conductance have the same order of magnitude as the experimental data of Reed et al. [1]. The calculated current at $V_{bias}$=1.5Volts is 0.050 μA, which is about 1.7 times that of Reed et al. of ~0.03μA. In comparison, previous calculations obtained currents of ~10 μA [33], ~10 μA (for BDT on jellium surface) [24], ~0.3 μA (for BDT-Au-atom on jellium surface) [24], ~20 μA [12], ~10μA [38], and ~10 μA [39] at the same bias voltage. The calculated differential conductance at $V_{bias}$=0 is 0.0062 μS, which is three times that of Reed et al. of ~0.002μS [1]. In comparison, previous calculations obtained $V_{bias}$=0 differential conductance of 5.0 μS [33], 3μS (for BDT on jellium surface) [24], ~0 μS (for BDT-Au-atom on jellium surface) [24], 4.8μS [12], 12μS [37], 7μS [38], and 6μS [39]. Here, it is emphasized that these previous calculations, except Ref. 24, considered a contact geometry, in which the S end of BDT is bonded with three Au atoms at the three-fold site of a flat Au(111) surface. The present study suggests that in the molecular-junction sample fabricated by Reed et al. [1] the S end of BDT is bonded with a single Au atom rather than multi Au atoms. Fig. 5(b) shows that the first peak in the C-V curve obtained in this study is located at 0.80 Volts, which is smaller than the experimental value obtained by Reed et al. [1] of ~1.3 Volts by 0.5 Volts. In comparison, the first peak in the C-V curve obtained by Di Ventra et al. was 2.5 Volts [24], which was larger than the experimental value by 1.2 Volts.

The current calculations have also been done by treating 5d electrons of the Au atoms in each slab, except the three Au atoms coordinated with the single Au atom bonded with the S atom, as core electrons. In this case, only delocalized Au 6sp electrons participate in the bonding among Au atoms and contribute to current. The calculated I-V characteristic is plotted in Fig. 6. This I-V curve differs from that shown in Fig. 5(a) in both shape and magnitude. At $V_{bias}$=1.0 Volt, the current is 0.0545μA, which is 1.7 times larger than 0.0318μA when

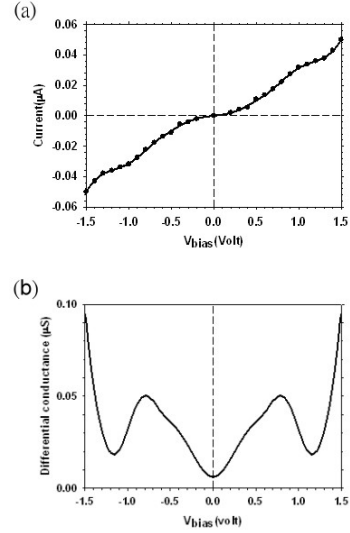

Fig. 5 (a) I-V characteristic and (b) Differential C-V characteristic of the gold-SAM-gold tri-layer system. In (a), the solid curve is obtained by fitting calculated values denoted by filled circles to 7-parameter rational functions. The circles in the negative bias voltage region are copied from those in the positive region by symmetry. The curve in (b) is obtained from the fitted functions.

all Au 5d electrons are treated as valence electrons. If 5d electrons of all Au atoms are treated as core electrons, the current at $V_{bias}$=1.0Volt is overestimated further to become 0.087μA, which is close to that obtained by Di Ventra et al. [24] of ~0.085μA when they considered that the S end of BDT is bonded with a single Au atom which is then attached to the flat jellium surface. The enhancement of current by confining Au 5d electrons in the ion cores can be understood by the following reasons. First, it is stated previously that Au 5d orbitals contribute substantially to the Au-S covalent bonding. The treatment of Au 5d states as core states deprives hybridization between Au 5d and S 3p orbitals, which underestimates the Au-S covalent bonding and consequently the potential barrier at the Au-BDT contact. Second, when Au 5d states are treated as valence states, part of Au 5d states are delocalized, so that the PDOS of Au 5d states at $E_F$ is approximately equal to those of 6s and 6p states as stated previously. Let $D_{Au}(E_F)$ be the total density of Au states at $E_F$, the PDOS of combined 6s and 6p states will be roughly $\frac{2}{3}D_{Au}(E_F)$. Assume that the current is contributed dominantly by delocalized 6sp states because they are much more itinerant than delocalized 5d states, then $I \propto \frac{2}{3}D_{Au}(E_F)$. If $\frac{1}{3}D_{Au}(E_F)$ of Au 5d states are replaced by Au 6sp states, then $I \propto D_{Au}(E_F)$, i.e. the current will be enhanced by 3/2=1.5 times, which is close to the calculated 1.7 times stated previously. This analysis suggests that the enhancement of the current be due to replacement of near-$E_F$ delocalized Au 5d states by much more itinerant delocalized 6sp states. Figure 3(a) shows that PDOS's of Au 6s and 6p states in bulk gold crystal and the three-layer slab are not free-electron-like, which is proportional to the square root of the energy measured from the bottom of the 6sp bands. The present result implies that the use of free-electron-like electronic states, which are more sp like than d like, and free-electron-like density of states to represent the electronic structure of the gold electrode may overestimate the current.

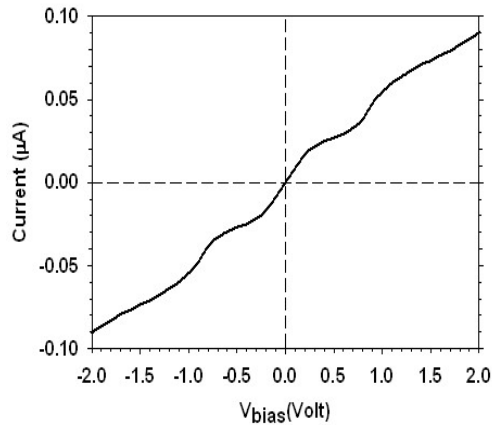

Fig. 6 I-V characteristic of the gold-SAM-gold tri-layer system when 5d electrons of the Au atoms in the slabs, except the three Au atoms coordinated with the single Au atom bonded with the S atom, are treated as core electrons.

**VI. Conclusion**

An integrated piecewise thermal equilibrium approach based on the first-principles calculation method has been developed to calculate I-V and differential C-V characteristics of the gold-BDT-gold molecular junction. The calculated currents and differential conductance have the same order of magnitude as experimental ones. Much better quantitative agreement between calculated currents and experimental data of Reed et al. suggests that the S end of BDT is bonded with a single Au atom rather than multi Au atoms considered in most previous theoretical studies. This study also finds a bias induced electron transfer between the two electrodes, which causes the local electronic structures in the two electrodes to be different. When the 5d electrons of all Au atoms are treated as core electrons, the current is overestimated, which can be understood as an underestimate of the Au-S covalent bonding and the potential barrier and the replacement of delocalized Au 5d carriers by more itinerant and longer ranged delocalized Au 6sp carriers in the electrodes.

**Acknowledgement:**
This work was supported by the National Science Council of Taiwan (contract number NSC 94-2112-M-110-012).